\documentclass[prl,longbibliography,twocolumn,nopacs,preprintnumbers,notitlepage,amsmath,amssymb,superscriptaddress]{revtex4-2}

\usepackage{amsmath}
\usepackage{amssymb}
\usepackage{amsbsy}
\usepackage{amsfonts}
\usepackage{graphicx}
\usepackage{esint}
\usepackage{dsfont}
\usepackage{xcolor}
\usepackage{mathrsfs}
\usepackage{enumerate}
\usepackage{mathtools}
\usepackage{hyperref}
\usepackage{pgffor}

\newcommand{\dd}[0]{\mathrm{d}}

\newcommand{\insertpdfpage}[2]{%
  \onecolumngrid           
  \clearpage
  \pagestyle{empty}         
  \thispagestyle{empty}     
  \null
  \vspace*{-1in}            
  \hspace*{-0.90in}         
  \makebox[\paperwidth][l]{%
    \includegraphics[page=#2,width=\paperwidth,height=\paperheight]{#1}%
  }%
  \clearpage
  \twocolumngrid           
  \pagestyle{plain}        
}

\begin{document}

\title{Beyond the Arcsine Law: Exact Two-Time Statistics of the Occupation Time in Jump Processes}

\author{Arthur Plaud}
\affiliation{Sorbonne Universit\'e, CNRS, Laboratoire de Physique Th\'eorique de la Mati\`ere Condens\'ee (LPTMC), 4 Place Jussieu, 75005 Paris, France}

\
\author{Olivier B\'enichou}
\affiliation{Sorbonne Universit\'e, CNRS, Laboratoire de Physique Th\'eorique de la Mati\`ere Condens\'ee (LPTMC), 4 Place Jussieu, 75005 Paris, France}

\begin{abstract}
Occupation times quantify how long a stochastic process remains in a region, and their single-time statistics are famously given by the arcsine law for Brownian and Lévy processes. By contrast, two-time occupation statistics—which directly probe temporal correlations and aging—have resisted exact characterization beyond renewal processes. In this Letter we derive exact results for generic one-dimensional jump processes, a central framework for intermittent and  discretely sampled dynamics. Using generalized Wiener–Hopf methods, we obtain the joint distribution of occupation time and position, the aged occupation-time law, and the autocorrelation function. In the continuous-time scaling limit, universal features emerge that depend only on the tail of the jump distribution, providing a starting point for exploring aging transport  in complex environments.
\end{abstract}

\maketitle

The occupation time, defined as the duration a system spends in a given state or region of space, is a fundamental observable in stochastic processes, with broad applications across physics, biology, and finance. The celebrated arcsine law, first uncovered by Lévy \cite{levy_sur_1940}, gives the distribution of the time $T_t$ spent by a one-dimensional standard Brownian motion on the positive side between $0$ and $t$:
\begin{equation}
\mathbb{P}(T_t = s) = \dfrac{1}{\pi \sqrt{s(t-s)}}
\label{eq:arcsine_law}
\end{equation}
Beyond Brownian motion, occupation time plays a central role in systems ranging from blinking quantum dots \cite{margolin_nonergodicity_2005} and spin glasses \cite{majumdar_exact_2002} to financial models \cite{cai_occupation_2010,guerin_joint_2016}, where it serves as a probe of ergodicity breaking and nonequilibrium dynamics. Since Lévy’s result, considerable effort has been devoted to computing the occupation-time distribution for stochastic processes, including Brownian motion with drift \cite{takacs_generalization_1996,godreche_statistics_2001} and absorbed \cite{randon-furling_residence_2018}, in higher dimensions \cite{barlow_extension_1989,desbois_occupation_2007}, diffusion in disordered media \cite{majumdar_local_2002}, active diffusion \cite{singh_generalised_2019,mukherjee_large_2024}, many-particle diffusion \cite{agranov_occupation_2019,burenev_occupation_2024}, continuous-time random walks \cite{bel_weak_2005,mendez_occupation-time_2025}, space-dependent diffusion \cite{del_vecchio_del_vecchio_generalized_2025}, random acceleration processes \cite{ouandji_boutcheng_occupation_2016}, and fractional Brownian motion \cite{sadhu_generalized_2018}.

The case of jump processes $\{X_n\}$, defined as discrete-time one-dimensional random walks via $X_{n+1} = X_n + \eta_n$, where the increments $\{\eta_n\}$ are independent and identically distributed, has also been widely studied. These processes play a central role in modeling stochastic dynamics \cite{majumdar_universal_2010,klinger_splitting_2022,klinger_extreme_2024,vezzani_fast_2024}: they (i) capture trajectories with intermittent or randomly reorienting ballistic motion, as observed in light scattering \cite{baudouin_signatures_2014,araujo_levy_2021} or self-propelled particles \cite{romanczuk_active_2012,solon_active_2015}; and (ii) reflect the fact that experimental time series are discretized by finite sampling. As a result, any observable extracted from data is inherently defined in discrete time and cannot be directly inferred from continuous-time models alone; in this discrete-time setting, the limiting distribution of the occupation time was obtained by Spitzer \cite{spitzer_combinatorial_1956}, who showed that it is universal for symmetric jump distributions $p(\eta)$, including symmetric Lévy flights, and coincides with the arcsine law \eqref{eq:arcsine_law}.

Despite these advances, previous studies have been essentially limited to \emph{single-time} observables. While informative, these quantities are time-local and cannot reveal the temporal correlations and history dependence that characterize aging nonequilibrium systems. Two-time observables directly probe this temporal structure. This raises a basic question: what becomes of L\'evy’s arcsine law when the process is \emph{aged}—that is, when the walk is allowed to evolve for \(n\) steps before the occupation is measured over a later window? 

For renewal systems—two-state processes $\sigma_t=\pm1$ where successive intervals between state changes are i.i.d.—this question has an essentially complete answer: Godr\`eche and Luck \cite{godreche_statistics_2000} computed the two-time correlator of the occupation time, and Akimoto \emph{et al.} \cite{akimoto_aging_2020} derived the aged occupation-time distribution $\mathbb{P}(T_{t+t'}-T_t=s)$.

\begin{figure}[h]
    \centering
    \includegraphics[width = 0.5\textwidth-9.0pt]{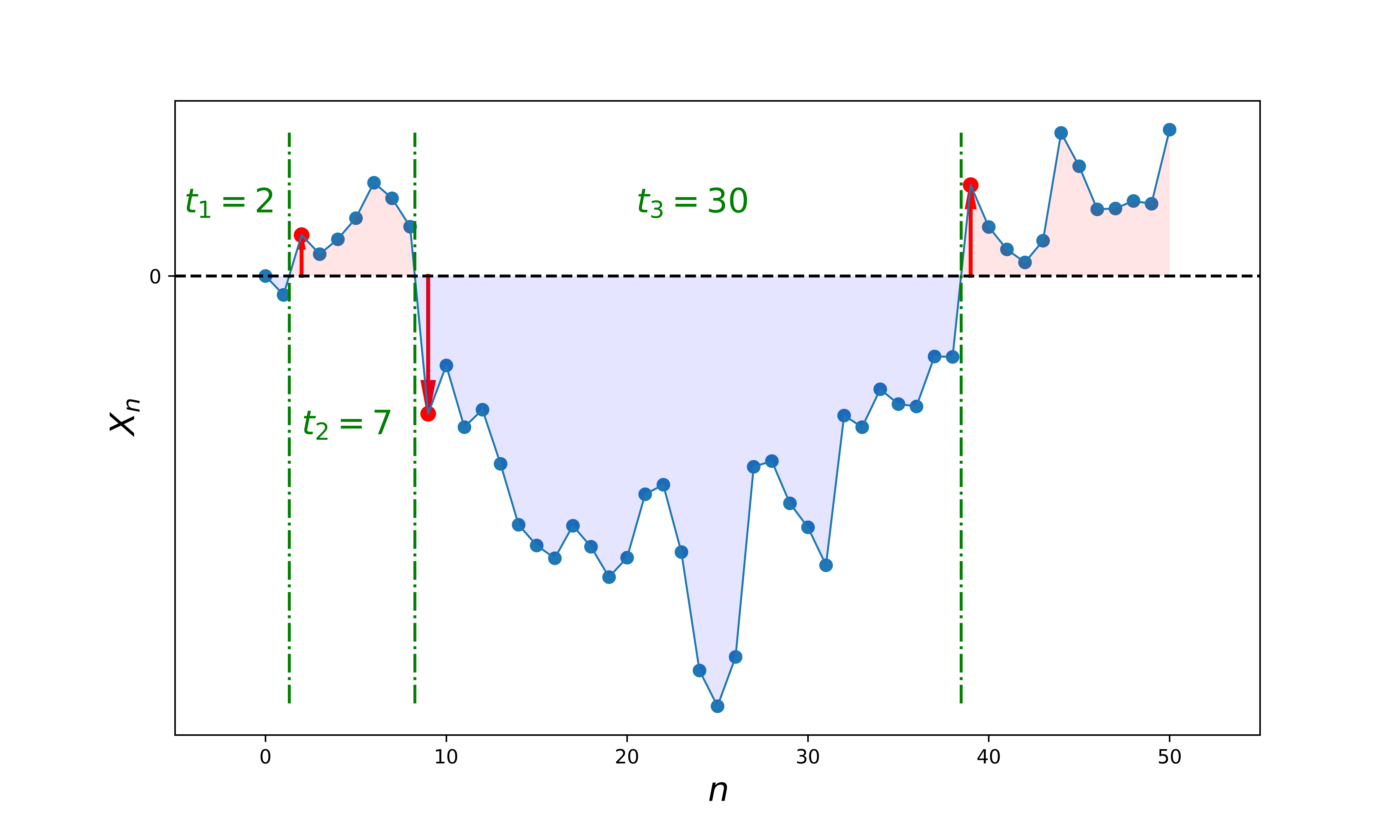}
    \caption{A discrete-time random walk \(X_n\) (i.e., a jump process) starting at \(X_0=0\). The first three \emph{excursions}—time intervals between consecutive sign changes of \(X_n\)—have durations \(t_i\). Segments with \(X_n<0\) are shown in blue and those with \(X_n>0\) in red. Each excursion ends with a jump that crosses the origin; the nonzero landing position defines the \emph{overshoot} (red arrows). Large excursions typically end with large overshoots, which restart the next excursion farther from \(0\) and tend to lengthen it. These overshoot-induced dependencies couple successive \(t_i\), so the sequence of excursions is \emph{not} renewal (durations are not i.i.d.).}
    \label{Illu Overshoot}
\end{figure}

However, renewal processes can only model situations where the trajectory decomposes into statistically independent time intervals. In the context of occupation time, these intervals correspond to excursions from zero—segments between successive zero-crossings. For jump processes, this decomposition fails: overshoots of the origin \cite{koren_leapover_2007,godreche_first_2025} introduce correlations between excursions, violating renewal assumptions (see Fig.~\ref{Illu Overshoot}).
Understanding how these correlations shape two-time occupation statistics is the main goal of this Letter. We overcome a key challenge: obtaining exact analytical results for two-time observables in non-renewal stochastic processes where excursion durations are intrinsically correlated.

More precisely, we compute the two-time probability distribution of the occupation time for arbitrary jump processes and their continuous-time scaling limits. This provides access to the aged occupation-time distribution and the corresponding two-time correlation functions. Notably, the jump–process results depart from renewal predictions even as \(t\) (aging time) and \(t'\) (observation window) tend to infinity at fixed \(r=t'/t\): overshoot–induced correlations persist, yielding tail-dependent edge Dirac masses and a distinct regular part in the aged occupation–time law, together with a different long–time crossover of the occupation–time autocorrelation.
 Our approach is based on (i) the joint statistics of the occupation time and the endpoint of a jump process starting at zero, and (ii) the occupation-time distribution for arbitrary starting positions.
Both quantities are of intrinsic theoretical interest, beyond their role in constructing two-time observables. Importantly, the framework applies to all jump processes, including asymmetric cases.

\emph{Joint Statistics of Occupation Time and Position.}  
Our first objective is the joint distribution \( \varphi_n(x,t) \equiv \mathbb{P}(X_n= x, T_n = t) \) of the endpoint \( X_n \) and the occupation time \( \displaystyle T_n = \sum_{k=1}^{n} \mathds{1}_{X_k \geq 0} \)  for all jump processes starting from \( X_0 = 0 \).
Conditioning on the position at step $n$ gives
\begin{equation}
\varphi_{n+1}(x,t) = \int_{-\infty}^{+\infty} \dd x' p(x-x') \varphi_{n}(x',t-\mathds{1}_{x\geq 0})
\label{eq:evolution}
\end{equation}
where \( p(x) \) denotes the (not necessarily symmetric) jump distribution. Since the occupation time increases by 1 only when \( x \geq 0 \), the second argument of \( \varphi_n \) within the integral depends on the sign of \( x \). 
We introduce the generating function
\(
\displaystyle G(x,\kappa,\xi) = \sum_{n=0}^{+\infty} \sum_{t=0}^{n} \xi^n \kappa^t \varphi_{n}(x,t),
\)
which satisfies the piecewise linear integral equation:
\begin{equation}
G(x,\kappa,\xi) = \delta(x) + \xi \kappa^{\mathds{1}_{x\geq 0}} \int_{-\infty}^{+\infty}  \dd x' p(x-x') G(x',\kappa,\xi).
\label{eq:wiener_hopf}
\end{equation}
This is analogous to standard Wiener–Hopf equations \cite{spitzer_wiener-hopf_1957,ivanov_resolvent_1994}:
\begin{equation}
G_0^{\pm}(x,\xi) = \delta(x) + \xi \int_{0}^{+\infty} \dd x' p(\pm x\mp x') G_0^{\pm}(x',\xi),
\end{equation}
with \( G_0^\pm(x,\xi) \) defined for \( x \geq 0 \) and equal to the generating functions of the semi-infinite propagators:
\begin{equation}
    G_0^{\pm}(x,\xi) = \sum_{n=1}^{\infty} \xi^n \mathbb{P}(X_0 = 0, X_{1, \dots, n-1} \in \mathbb{R}^{\pm}, X_n = \pm x).
\end{equation}
Equation~\eqref{eq:wiener_hopf} is thus a generalized Wiener–Hopf equation. Its solution reads
\begin{multline}
\begin{cases}
G(x,\kappa,\xi) = \int_0^{+\infty} \dd x' G_0^-(x',\xi)G_0^+(x+x',\xi \kappa)\\
G(-x,\kappa,\xi) = \int_0^{+\infty} \dd x' G_0^+(x',\xi \kappa)G_0^-(x+x',\xi ),
\end{cases}
\label{eq:realspace_G}
\end{multline}
and, using the Laplace transforms of \( G_0^\pm \) \cite{mounaix_asymptotics_2018}, its Fourier transform:
\begin{multline}
\int_{-\infty}^{+\infty} \dd x e^{i s x} G(x,\kappa,\xi)
= \dfrac{1}{\sqrt{(1-\xi \tilde{p}(s))(1-\xi \kappa \tilde{p}(s))}} \\
\times \exp\left[\dfrac{i }{2 \pi}\fint_{-\infty}^{+\infty} \dfrac{\dd k}{k-s} \log\left(\dfrac{1- \xi \kappa \tilde{p}(k)}{1- \xi \tilde{p}(k)}\right)\right]
\label{eq:fourier_G}
\end{multline}
where \( \tilde{p}(k) = \int_{\mathbb{R}} \dd x e^{i k x} p(x) \), the integral being taken in the principal value sense.

This general expression calls for several remarks: (i) Similar joint statistics \cite{wu_occupation_2017} have recently been derived in the continuous-time setting of Lévy processes. In contrast, Eq.~\eqref{eq:fourier_G} (a) follows from elementary steps, (b) depends only on \(p\), and (c) is convenient for two-time observables. (ii) Known marginals are recovered (see SM): \(\kappa=1\) yields the law of \(X_n\); \(s=0\) yields the occupation-time distribution (discrete arcsine law for symmetric processes). (iii) Despite explicit dependence on \(p\), the correlation between endpoint sign and occupation time is universal. For symmetric processes,
\begin{equation}
\sum_{n=0}^{\infty}\xi^n \sum_{t=0}^{n}\kappa^t\, \varphi_n^+(t)
=
\frac{1}{2}
+
\frac{\xi\kappa}{\,1-\xi\kappa+\sqrt{(1-\xi)(1-\xi\kappa)}\,},
\label{eq:sign_occ}
\end{equation}
where \( \varphi_n^+(t) = \mathbb{P}(T_n = t,X_n \geq 0)\) ,and with no dependence on the jump distribution. Even for large \(n\) the endpoint sign strongly constrains the occupation-time distribution (the asymmetric case is in SM). (iv) Equation~\eqref{eq:fourier_G} is well suited for asymptotic analysis.
\begin{equation}
\tilde{p}(k)\underset{k\to 0}{=} 1-(C|k|)^{\alpha}\bigl(1-i\tilde{\beta}\,\mathrm{sgn}\,k\bigr)+o(|k|^{\alpha}),
\label{eq:smallk_ptilde}
\end{equation}
implies convergence of the jump process to a stable process \cite{kyprianou_fluctuations_2014} of index \(\alpha\) and asymmetry \( \tilde{\beta} \) \footnote{Here, we did not use the classic parameters defining stable processes, mainly for simplicity. To recover the usual parametrization, use \( \tilde{\beta} = \beta \tan\left(\frac{\pi \alpha}{2}\right) \) when \( \alpha \neq 1 \) and \( \tilde{\beta} = \mu \) when \( \alpha =1 \) }. Analyzing \eqref{eq:fourier_G} in the scaling regime yields the joint law for continuous stable processes.

To obtain the two-time occupation-time distribution \( \mathbb{P}(T_n = t,T_{n+n'} = t+t') \), we also need \( \varphi_n(\bullet,t|x) \equiv \mathbb{P}(T_n = t|X_0 = x) \).
Indeed, using the Markov property and integrating over all possible \( x \) at time \( n \),
\begin{equation}
    \mathbb{P}(T_n = t,T_{n+n'} = t+t') = \int_{-\infty}^{+\infty} \dd x \, \varphi_n(x,t) \, \varphi_{n'}(\bullet,t'|x).
    \label{eq:two_time}
\end{equation}
\( \varphi_n(\bullet,t|x) \) is derived using similar methods as for the joint distribution \( \varphi_n(x,t) \). Partitioning over the first step gives
\begin{equation}
\varphi_{n+1}(\bullet,t|x) = \int_{-\infty}^{+\infty} \dd x' p(x'-x) \varphi_{n}(\bullet,t-\mathds{1}_{x' \geq 0}|x'),
\end{equation}
and the generating function
\( \displaystyle
G(\bullet,\kappa,\xi \vert x) = \sum_{n=0}^{+\infty} \sum_{t=0}^{n} \xi^n \kappa^t \varphi_{n}(\bullet,t\vert x)
\)
satisfies:
\begin{equation}
G(\bullet,\kappa,\xi|x) = 1 + \xi \int_{-\infty}^{+\infty} \dd x' \kappa^{\mathds{1}_{x'\geq 0}} p(x'-x) G(\bullet,\kappa,\xi|x').
\label{eq:wiener_hopf_2}
\end{equation}
Mapping its derivative with respect to \( x \) to \( G(x,\kappa,\xi) \) in Eq.~\eqref{eq:realspace_G} yields:
\begin{multline}
G(\bullet,\kappa,\xi|\pm x) = \dfrac{1}{1-\xi\kappa^{\frac{1\pm 1}{2}}}\\ 
+ \left(\kappa^{\mp 1}-1\right) G(\bullet,\kappa,\xi\vert 0) \int_x^{+\infty} \dd x' G(\mp x',\frac{1}{\kappa},\xi \kappa).
\label{eq:realspace_Gbullet}
\end{multline}
Collecting these results, we now have  representations for both \(G(x,\kappa,\xi)\) and \(G(\bullet,\kappa,\xi\,|\,x)\). Eqs.~\eqref{eq:fourier_G} and \eqref{eq:realspace_Gbullet} not only stand as independent results characterizing jump-process dynamics, but also—as shown below—provide full access to two-time occupation-time statistics.

\emph{Aged occupation-time distribution.}
As a first two-time observable, we focus on the aged distribution \( \mathbb{P}(T_{n+n'} - T_n = t) \), which probes non-stationary dynamics.
Using Eq.~\eqref{eq:two_time}, we obtain the triple generating function \( \displaystyle \hat{G}_{\text{aged}}(\xi_1,\xi_2,\kappa) = \sum_{n,n',t'=0}^{+\infty} \xi_1^n \xi_2^{n'} \kappa^{t'} \mathbb{P}(T_{n+n'}-T_n = t') \) as:
\begin{multline}
\hat{G}_{\text{aged}}(\xi_1,\xi_2,\kappa) = \int_{-\infty}^{+\infty} \dd x \, G(x,1,\xi_1)\, G(\bullet,\kappa,\xi_2|x).
\label{eq:GF_aged}
\end{multline}
This triple generating function gives access to the full discrete distribution, including short-time dynamics.We can process further in the continuous scaling regime. In the scaling limit \( n, n', t \to \infty \) with \( t/n' \sim 1 \) and \( n/n' \sim 1 \),
\begin{equation}
\mathbb{P}(T_{n+n'}-T_n = t) \sim \frac{1}{n'}f\!\left(\frac{t}{n'},\frac{n}{n'}\right),
\label{scaling form aged distribution}
\end{equation}
corresponding to \( \xi_1,\xi_2,\kappa \to 1 \) at fixed \( \lambda = \frac{1-\xi_2}{1-\xi_1} \) and \( \mu = \frac{1-\kappa}{1-\xi_1} \). Evaluating Eq.~\eqref{eq:GF_aged} at leading order in the regime using Eqs.~\eqref{eq:fourier_G} and \eqref{eq:realspace_Gbullet} yields an integral equation for \( f(s,r) \) depending only on the small-\(k\) behavior \eqref{eq:smallk_ptilde}. For clarity we restrict to \( \tilde{\beta}=0 \) (asymmetry in SM):
\begin{multline}
\int_0^{+\infty} \dd r \int_0^{+\infty} \dd s \, \dfrac{f(s,r)}{(r+\lambda+s \mu)^2} = \dfrac{2\lambda+\mu}{2\lambda(\lambda+\mu)}\\
   + \dfrac{\mu}{\pi \sqrt{\lambda(\lambda+\mu)}}  \int_0^{+\infty} \dfrac{\dd k}{k(1 +k^\alpha)}\dfrac{1}{\sqrt{[\lambda+\mu+k^\alpha][\lambda+k^\alpha]}}\\
   \times \sin\left[\dfrac{k}{\pi}\fint_{0}^{+\infty} \dfrac{\dd s}{s^2-k^2} \log\left(\dfrac{\lambda + \mu +s^\alpha}{\lambda+ s^\alpha }\right) \right].
\label{eq:integral_aged}
\end{multline}
Equation~\eqref{eq:integral_aged} is a cornerstone of this work. It fully characterizes the scaling function \( f(s,r) \) governing aging. It generalizes the classical arcsine law \eqref{eq:arcsine_law} to capture temporal structure induced by aging and heavy-tailed dynamics. Known analytical results are recovered in the Brownian case \( \alpha = 2 \); in other cases the equation is solved numerically. It also provides direct access to key observables—such as singular contributions, the forward recurrence time \( F_n \) (first crossing of \(0\) after time \( n \)), and moments. In this sense, Eq.~\eqref{eq:integral_aged} establishes a framework for aging phenomena beyond renewal stochastic dynamics, extending these results to the much broader setting of jump processes with correlated excursions. 
\begin{figure}[h]
    \centering
    \includegraphics[width=\textwidth/2 +12.0 pt]{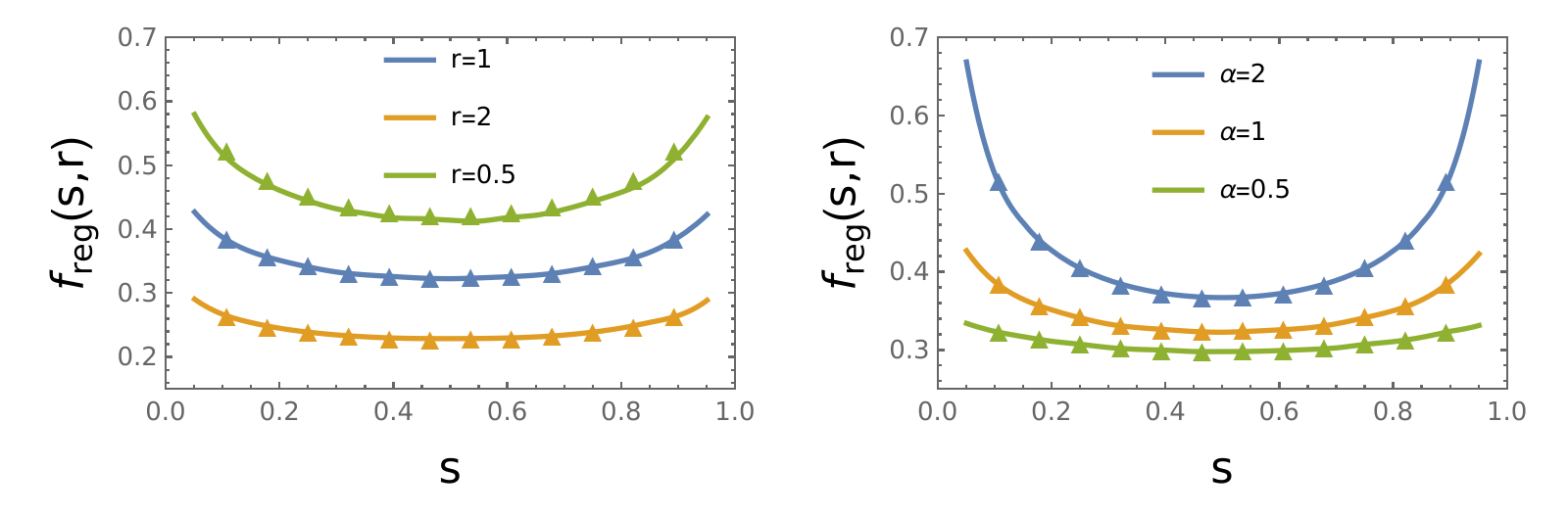}
    \caption{The regular part  \( f_{\text{reg}}(s,r) \) of the limiting distribution of the aged occupation time, obtained by numerically solving \eqref{eq:integral_aged}. On the left, the distribution is shown in the Cauchy case \( \alpha =1 \) for \( 3 \) different values of the aging ratio \( r \). On the right this time, the aging ratio is fixed to \( 1 \) but \( 3 \) different universality classes are presented. Agreement with numerical simulations (triangles) is excellent.}
    \label{plot aged distribution}
\end{figure}

Compared with the unaged arcsine law, aging modifies the occupation-time law in two specific ways: it creates Dirac peaks at \(s=0\) and \(s=1\) and reshapes the edge behavior of the regular part with \(\alpha\)-dependent exponents. As soon as \(r>0\), the process can remain on one side of the origin for the entire interval \([n,n+n']\) with non-zero probability—even in the large-time limit; this persistence is precisely what generates the Dirac peaks at \(s=0\) and \(s=1\) in \(f(s,r)\).
This leads to the decomposition
\begin{equation}
    f(s,r) = q(r)\, \bigl[\delta(s)+\delta(1-s)\bigr] + f_{\text{reg}}(s,r),
\end{equation}
where \(q(r)>0\) for \(r>0\), and \(f_{\text{reg}}\) is normalized to \(1-2q(r)\).
This decomposition already appears in the Brownian case: Akimoto \emph{et al.}~\cite{akimoto_aging_2020} obtained explicit forms for \(q(r)\) and \(f_{\text{reg}}(s,r)\).
Beyond Brownian motion (\(\alpha=2\)), however, the situation is qualitatively different and this result provides little information for \( 0<\alpha < 2 \). The unaged limit \(r=0\) recovers the arcsine law; at the opposite extreme \(r\to\infty\) the distribution becomes purely singular, with \(q(r)\to\frac12\) and \(f_{\text{reg}}\to 0\). In between, both the singular weight and the shape of the regular part are \(\alpha\)-dependent, delineating distinct universality classes. Below we determine \(q(r)\) exactly and characterize \(f_{\text{reg}}(s,r)\)—including its edge behavior—across the full range \(0<\alpha\leq2\).

To compute \(q(r)\), we consider the limit \(\mu\to\infty\) in Eq.~\eqref{eq:integral_aged}, which isolates the singular contribution. This leads to an exact expression for \(q(r)\). 
Using \( q(r)=\lim_{n\to\infty}\frac{1}{2}\,\mathbb{P}\!\left(\frac{F_n}{n}\ge \frac{1}{r}\right) \),
we obtain the asymptotic forward-recurrence-time distribution \( f_{\text{FRT}}(r) = \underset{n\rightarrow +\infty}{\lim} \mathbb{P}\left(\dfrac{F_n}{n} = r\right) \) as:
\begin{multline}
    f_{\text{FRT}}(r) = \dfrac{2}{\pi \alpha r \sqrt{1+r}} \sin\left[ \dfrac{r^{\frac{1}{\alpha}}}{\pi} \fint_0^{+\infty} \dd k \, \dfrac{\log(1+k^\alpha)}{k^2 -r^{\frac{2}{\alpha}}}\right].
    \label{FRT distribution}
\end{multline}
This generalizes the aged first-passage-time concept (Godrèche–Luck \cite{godreche_statistics_2001}, for renewal processes) to jump processes and captures the statistics of the first crossing of \(0\) after time \(n\).

\begin{figure}[h]
    \centering
    \includegraphics[width=\textwidth/2 - 32.0 pt]{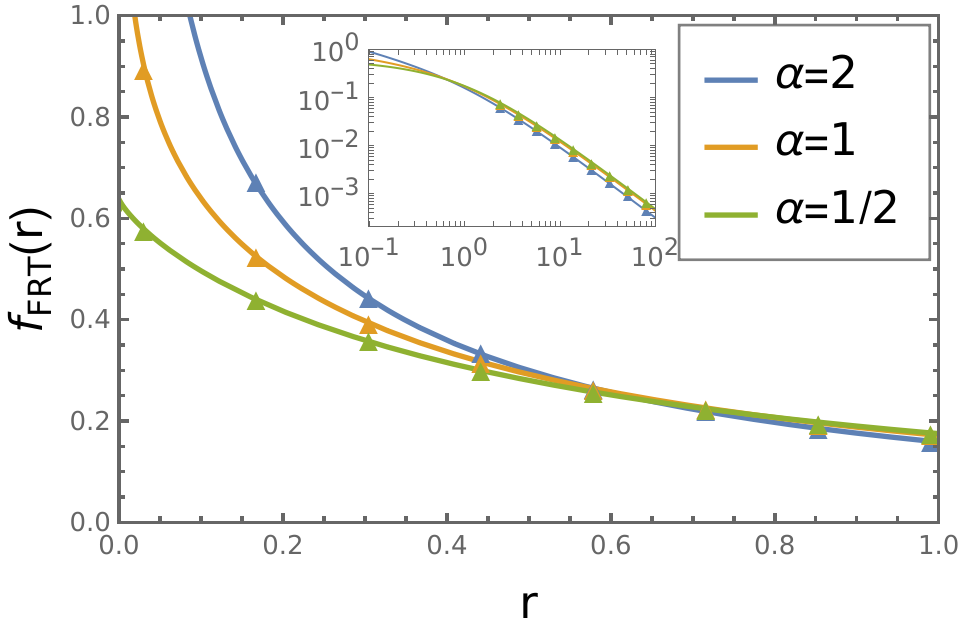}
    \caption{The limiting distribution of the rescaled forward recurrence time \( F_n/n \), for three values of \( \alpha \). For large \(u\), \(f_{\text{FRT}}(u)\sim \frac{2}{\pi \alpha}\sin\!\bigl(\frac{\pi \alpha}{4}\bigr)\,u^{-3/2}\), consistent with the universal Sparre–Andersen prediction. The small-\(u\) behavior diverges for \( \alpha \geq 1 \) and remains finite for \( \alpha < 1 \). Agreement with numerical simulations (triangles) is excellent.}
    \label{plot FRT distribution}
\end{figure}

To further characterize aging effects, we now focus on the behavior of the regular part \( f_{\text{reg}}(s,r) \) near the edges \( s = 0 \) and \( s = 1 \).
Whereas the Dirac peaks originate from trajectories that never cross the origin, this edge behavior encodes how likely the process is to cross the origin while remaining almost entirely on the same side.
In the Brownian case (\( \alpha = 2 \)), the inverse square-root divergence of the arcsine law persists: aging affects the prefactor but not the type of divergence.
However, this picture changes drastically for \( \alpha < 2 \). Depending on the universality class, the divergence softens or disappears entirely:
\begin{equation}
    \begin{cases}
    1 < \alpha \leq 2 : f_{\text{reg}}(s,r) \underset{s \rightarrow 0}{\sim} \; D_\alpha(r) s^{\frac{1-\alpha}{\alpha}} \\
    \alpha = 1 : f_{\text{reg}}(s,r) \underset{s \rightarrow 0}{\sim} \; D_1(r) \log\left(\dfrac{1}{s}\right) \\
    0 < \alpha < 1 : f_{\text{reg}}(s,r) \underset{s \rightarrow 0}{\sim} \; D_\alpha(r),
    \end{cases}
    \label{eq:low_s_edge}
\end{equation}
revealing a sharp crossover at \( \alpha = 1 \). For \( \alpha \geq 1 \) the regular part still diverges (more weakly than arcsine); for \( \alpha < 1 \), \( f_{\text{reg}} \) does not diverge at the edges. Interestingly, these changes occur for arbitrarily small values of \( r > 0 \): the aging-induced regularization of the edges is immediate, as soon as the system is no longer observed from its initial time. The prefactors \( D_\alpha(r) \) can be obtained analytically as shown in SM.

To track how \( f(s,r) \) evolves with \( r \), consider its moments. For symmetric processes, \( f(s,r) = f(1-s,r) \), so the first moment \(\int_0^1 \dd s f(s,r) \) equals \( \frac{1}{2} \) for all values of \( r \). Concentration of the probability near \( s=0,1 \) as \( r \) increases is captured by the second moment
\(
F_2(r) = \int_0^1 \dd s \, s^2 f(s,r),
\)
which increases from \( F_2(0)=3/8 \) (arcsine) to \( F_2(\infty)=1/2 \) (purely singular). Differentiating Eq.~\eqref{eq:integral_aged} twice with respect to \( \mu \) at \( \mu=0 \) yields an integral equation for \(F_2(r)\); its solution is:
\begin{multline}
F_2(r) = \dfrac{1}{2} +\int_0^{\frac{1}{r}} \dfrac{\dd q}{2\pi^2 \alpha q}\dfrac{(1-rq)^2}{1+q} \fint_{-\infty}^{+\infty} \dfrac{\dd v}{v-q^{\frac{1}{\alpha}}}\dfrac{1}{1+\vert v \vert^\alpha}.
\label{eq:second_moment}
\end{multline}
The limits \( r \to 0 \) and \( r \to \infty \) are universal, but elsewhere \( F_2(r) \)  depends continuously on \( \alpha \).
Notably, convergence to the singular regime is faster for smaller \( \alpha \), reflecting weaker memory effects in processes with heavy-tailed increments.

\emph{Autocorrelation of the Occupation Time.}
We now turn to the autocorrelation of the occupation time—arguably its most fundamental two-time observable.
This quantity probes how the system’s history influences future occupancy, and provides a direct measure of temporal correlations.
It is defined by
\(
C(n,n') = \langle T_n (T_{n+n'} - T_n) \rangle.
\)
The associated generating function \( \hat{C}(\xi_1,\xi_2) = \sum_{n,n'=0}^{+\infty} \xi_1^n \xi_2^{n'} C(n,n') \) satisfies:
\begin{multline}
\hat{C}(\xi_1,\xi_2) = \int_{-\infty}^{+\infty} \dd x \, \left.\dfrac{\partial G(x,\kappa,\xi_1)}{\partial \kappa}\right|_{\kappa=1} \,
\left.\dfrac{\partial G(\bullet,\kappa,\xi_2|x)}{\partial \kappa}\right|_{\kappa=1}.
\label{eq:gen_autocorr}
\end{multline}
In the scaling limit \( n, n' \to \infty \) with \( r = n/n' \) fixed,
\(
C(n,n') \sim n n' \, c(r),
\)
where \( c(r) \) depends only on \( \alpha \) and \( \tilde{\beta} \). For symmetric processes (general case in SM),
\begin{multline}
    c(r) = \dfrac{1}{4} + \int_0^{+\infty} \dd k\,\dfrac{r^2 - (r-k^\alpha)^2 \mathds{1}_{k^\alpha\leq r}}{4\pi^2 rk^{\alpha+1}(1+k^\alpha)} \fint_{-\infty}^{+\infty} \dfrac{\dd s}{k-s} \dfrac{1}{1+|s|^\alpha}.
    \label{eq:autocorr_scaling_function}
\end{multline}
This reveals a crossover between universal behaviors. The value \( c(+\infty) = \frac{3}{8} \) reflects the non-decaying correlation between occupation time and endpoint (extractable from \eqref{eq:sign_occ}). In contrast, \( r =0 \) decorrelates the two intervals (finite-range correlations between starting point and occupation time). The leading correction to \( c(0)\) for \( n' \gg n \) can be computed for all universality classes, including asymmetric ones, and gives the decay behavior of the autocovariance of the occupation time for all universality classes, in the regime \(n' \gg n \). If we denote \( A(n,n') = \left\langle \dfrac{T_n}{n}\dfrac{T_{n+n'} - T_n}{n'} \right\rangle  - \left\langle \dfrac{T_n}{n}\right\rangle \left\langle \dfrac{T_{n+n'} - T_n}{n'} \right\rangle \) the rescaled autocorrelation of the occupation time, we have in the regime $r=n/n'\to0$:
\begin{equation}
A(n,n') \sim
\begin{cases}
A^{1}_{\alpha,\tilde\beta}\,\, r^{1/\alpha}, & \alpha>1,\\[2pt]
\frac{\, r\,(\log r)^2}{4\pi^2\!\left(1+\tilde\beta^2\right)}, & \alpha=1,\\[2pt]
-\, A^{2}_{\alpha,\tilde\beta}\,\, r\, \log r, & \alpha<1,
\end{cases}
\label{eq:autocorr_decay}
\end{equation}
with explicit prefactors:
\begin{multline}
\begin{cases}A^1_{\alpha,\tilde{\beta}} = \dfrac{\csc\left(\frac{\pi}{\alpha}\right)^2}{\alpha^2 \Gamma\left(2-\frac{1}{\alpha}\right)\Gamma\left(2+\frac{1}{\alpha}\right)}\cos\left(\frac{\arctan(\tilde{\beta})}{\alpha}\right)^2 \\
A^2_{\alpha,\tilde{\beta}} = \dfrac{\tan\left(\frac{\pi \alpha}{2}\right) - \tilde{\beta}^2 \cot\left(\frac{\pi \alpha}{2}\right)}{4 \pi \alpha(1+\tilde{\beta}^2)}.
\end{cases}
\end{multline}
This matches the crossover in Eq.~\eqref{eq:low_s_edge}, distinguishing \( \alpha > 1 \) from \( \alpha < 1 \). Correlations remain long-ranged—a hallmark of nonequilibrium dynamics—and become more pronounced as \( \alpha \to 2 \).

\emph{Conclusion.}
We provided the first exact analytical framework to compute two-time occupation statistics for generic one-dimensional jump processes, beyond the renewal paradigm. Our results include: (i) the joint distribution of occupation time and position; (ii) the full aged distribution; and (iii) its two-time autocorrelation, obtained for arbitrary jump distributions—including asymmetric and heavy-tailed cases—via a generalized Wiener–Hopf approach. A central result is an explicit integral equation governing the scaling form of the aged occupation-time distribution generalizing the arcsine law, revealing Dirac peaks, nontrivial scaling functions, and \(\alpha\)-dependent edge regularization. We also derive the asymptotic forward-recurrence distribution and clarify autocorrelation scaling. This framework offers a starting point for systematic studies of temporal correlations of additive functionals \(A_n=\sum_{k=1}^{n} a(X_k)\) beyond renewal systems, with potential relevance to aging transport in complex environments.

\vspace{12pt}

\insertpdfpage{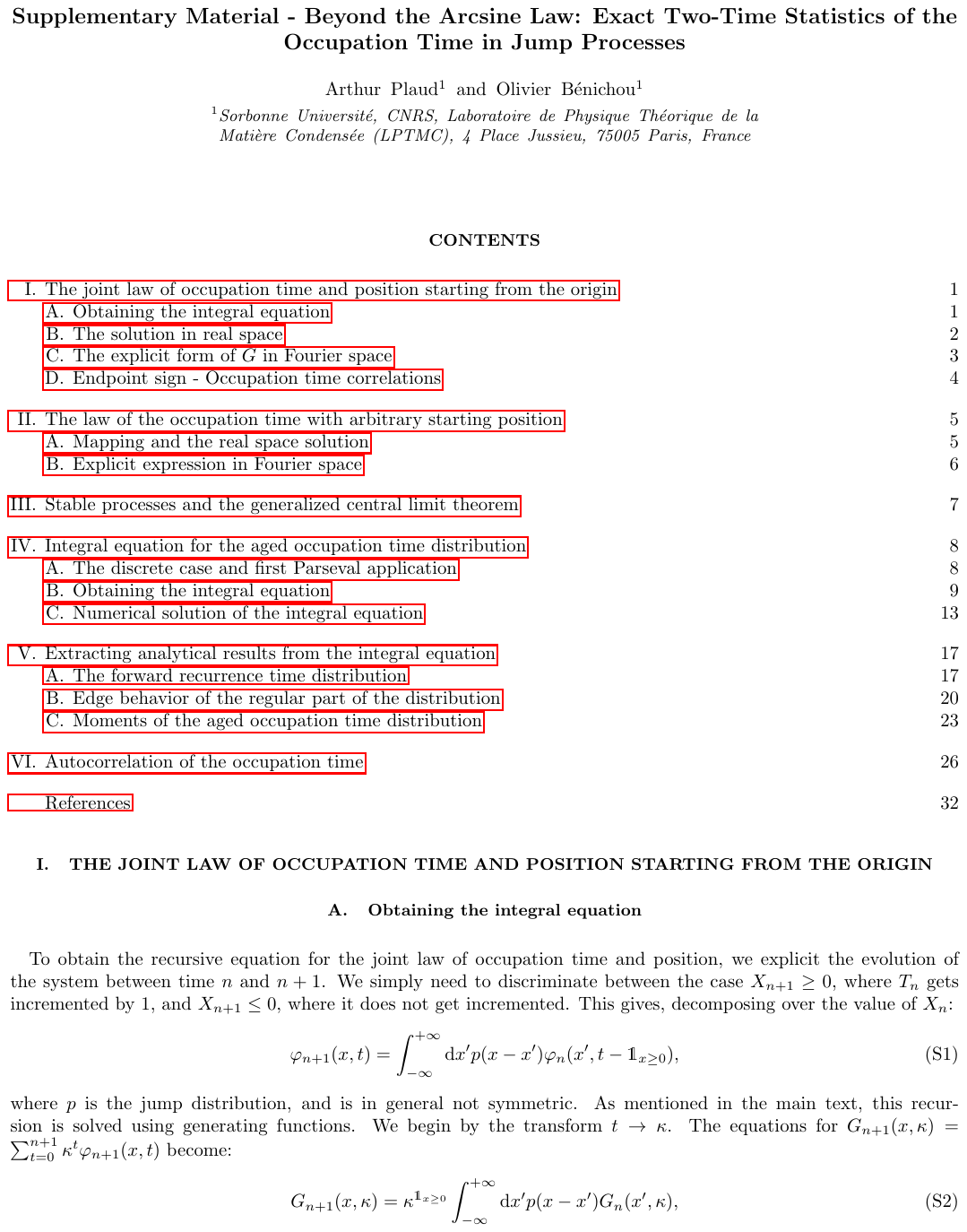}{1}
\insertpdfpage{Occupation__Time_Jump_Processes-14.pdf}{2}
\insertpdfpage{Occupation__Time_Jump_Processes-14.pdf}{3}
\insertpdfpage{Occupation__Time_Jump_Processes-14.pdf}{4}
\insertpdfpage{Occupation__Time_Jump_Processes-14.pdf}{5}
\insertpdfpage{Occupation__Time_Jump_Processes-14.pdf}{6}
\insertpdfpage{Occupation__Time_Jump_Processes-14.pdf}{7}
\insertpdfpage{Occupation__Time_Jump_Processes-14.pdf}{8}
\insertpdfpage{Occupation__Time_Jump_Processes-14.pdf}{9}
\insertpdfpage{Occupation__Time_Jump_Processes-14.pdf}{10}
\insertpdfpage{Occupation__Time_Jump_Processes-14.pdf}{11}
\insertpdfpage{Occupation__Time_Jump_Processes-14.pdf}{12}
\insertpdfpage{Occupation__Time_Jump_Processes-14.pdf}{13}
\insertpdfpage{Occupation__Time_Jump_Processes-14.pdf}{14}
\insertpdfpage{Occupation__Time_Jump_Processes-14.pdf}{15}
\insertpdfpage{Occupation__Time_Jump_Processes-14.pdf}{16}
\insertpdfpage{Occupation__Time_Jump_Processes-14.pdf}{17}
\insertpdfpage{Occupation__Time_Jump_Processes-14.pdf}{18}
\insertpdfpage{Occupation__Time_Jump_Processes-14.pdf}{19}
\insertpdfpage{Occupation__Time_Jump_Processes-14.pdf}{20}
\insertpdfpage{Occupation__Time_Jump_Processes-14.pdf}{21}
\insertpdfpage{Occupation__Time_Jump_Processes-14.pdf}{22}
\insertpdfpage{Occupation__Time_Jump_Processes-14.pdf}{23}
\insertpdfpage{Occupation__Time_Jump_Processes-14.pdf}{24}
\insertpdfpage{Occupation__Time_Jump_Processes-14.pdf}{25}
\insertpdfpage{Occupation__Time_Jump_Processes-14.pdf}{26}
\insertpdfpage{Occupation__Time_Jump_Processes-14.pdf}{27}
\insertpdfpage{Occupation__Time_Jump_Processes-14.pdf}{28}
\insertpdfpage{Occupation__Time_Jump_Processes-14.pdf}{29}
\insertpdfpage{Occupation__Time_Jump_Processes-14.pdf}{30}
\insertpdfpage{Occupation__Time_Jump_Processes-14.pdf}{31}
\insertpdfpage{Occupation__Time_Jump_Processes-14.pdf}{32}

\end{document}